\documentclass[11pt]{article}
\usepackage{newtxtext,newtxmath}
\usepackage{graphicx}
\usepackage[font=footnotesize,labelfont=bf]{caption}

\usepackage{titling}
\usepackage[letterpaper,margin=1in]{geometry}
\usepackage{authblk}

\makeatletter
\def\@maketitle{%
  \newpage
  \begin{center}%
  \let \footer \thanks
    {\LARGE \sffamily \bfseries \@title \par}%
    \vskip 1.5em%
    {\normalsize
      \lineskip .5em%
      \begin{tabular}[t]{c}%
        \@author
      \end{tabular}\par}%
    \vskip 1em%
    {\small \@date}%
  \end{center}%
  \par
  \vskip -4em}
\makeatother

\usepackage{titlesec}
\titleformat{\section}
  {\sffamily\Large\bfseries}
  {\thesection}
  {1em}
  {}

\titleformat{\subsection}
  {\bfseries}            
  {\thesubsection}
  {0.8em}
  {\sffamily}

\usepackage{abstract}

\usepackage[numbers,super,sort&compress]{natbib}

\makeatletter
\renewcommand{\fnum@figure}{\textbf{Figure \thefigure}}
\renewcommand{\fnum@table}{\textbf{Table \thetable}}
\makeatother
\usepackage[version=4]{mhchem}
\usepackage[utf8]{inputenc}
\usepackage[T1]{fontenc}
\usepackage[dvipsnames]{xcolor}
\usepackage[table]{xcolor} 
\usepackage{dcolumn}
\usepackage{bm}
\usepackage{longtable}
\usepackage[colorlinks=true, linkcolor=RoyalBlue, citecolor=RoyalBlue]{hyperref}
\usepackage[per-mode=symbol, separate-uncertainty]{siunitx}
\usepackage[capitalise]{cleveref}
\usepackage{enumerate}
\usepackage{lineno}

\usepackage{url}

\colorlet{highlight}{orange!70!black}
\colorlet{warning}{red!90!black}
\newif\ifshowhighlight
\showhighlighttrue        

\newcommand{\papertitle}{Microwave-to-optical transduction using magnon–exciton coupling in a layered antiferromagnet}

\title{\papertitle}
\author[1]{Pratap Chandra Adak\thanks{ padak@ccny.cuny.edu}}
\author[1]{Iris McDaniel}
\author[1]{Suvodeep Paul}
\author[1,7]{Caleb Heuvel-Horwitz}
\author[1]{Bikash Das}
\author[2]{Vitali Kozlov}
\author[3]{Kseniia~Mosina}
\author[4]{Arun Ramanathan}
\author[4]{Xavier Roy}
\author[3]{Zdeněk~Sofer}
\author[5]{Tian Zhong}
\author[6]{Akashdeep Kamra}
\author[2]{Arno Thielens}
\author[2,7,8]{Andrea Alù}
\author[1,7]{Vinod M. Menon\thanks{ vmenon@ccny.cuny.edu}}
\affil[1]{Department of Physics, City College of New York, New York, NY 10031, USA.}
\affil[2]{Photonics Initiative, CUNY Advanced Science Research Center, New York, NY 10031, USA.}
\affil[3]{Department of Inorganic Chemistry, University of Chemistry and Technology Prague, Prague,166 28 Czech Republic.}
\affil[4]{Department of Chemistry, Columbia University, New York, NY 10027, USA.}
\affil[5]{Pritzker School of Molecular Engineering, University of Chicago, Chicago, IL, USA.}
\affil[6]{Department of Physics and Research Center OPTIMAS, Rheinland-Pfälzische Technische Universität, Kaiserslautern-Landau, 67663 Kaiserslautern, Germany.}
\affil[7]{Department of Physics, Graduate Center of the City University of New York (CUNY), New York, NY 10016, USA.}
\affil[8]{Department of Electrical Engineering, The City College of New York, New York, NY 10031, USA.}

\begin{document}
\date{}

\maketitle
\begin{abstract}
{Coherent interfaces between microwave-frequency quantum systems and low-loss optical links are essential for quantum networks.
 However, existing microwave--optical transducers often trade conversion efficiency against added noise, bandwidth, and device integrability. 
 Here, we demonstrate coherent microwave-to-optical transduction based on magnon--exciton coupling in the layered antiferromagnet CrSBr. 
 Driving the antiferromagnetic resonance with microwave signals imprints coherent modulation on a reflected optical probe, generating optical sidebands that are resonantly enhanced near excitonic transitions.
 While prior magnon-based approaches to microwave-to-optical transduction have typically relied on intrinsically weak off-resonant magneto-optical effects (e.g., Faraday rotation), our scheme exploits strong light--matter interactions at exciton resonances. 
 Even in a bulk crystal without cavity enhancement, we observe coherent conversion over an intrinsically broadband window of ${\sim300}$~MHz.
 We further show that multiple exciton-polariton resonances inherit the magnon-coupled response, suggesting a route to broaden the usable optical detuning range and to mitigate optical dissipation.
 Our results establish magnon-coupled excitons in layered magnets as a scalable platform for broadband microwave--optical interfaces, with pathways to higher cooperativity via reduced magnetic volume and cavity integration.}  \hfill \break
 \end{abstract}
 
\section*{Introduction}

Quantum hardware platforms operate across disparate energy scales: superconducting qubits function in the microwave (GHz) domain, whereas systems such as trapped ions, neutral atoms, typically utilize the optical (THz) regime~\cite{kurizki_quantum_2015}, and photonic quantum technologies typically operate at telecom wavelengths. Realizing a functional quantum network requires interoperability between these diverse platforms.
For example, while superconducting circuits provide a leading architecture for quantum information processing, optical photons serve as the ideal "flying qubits" for long-distance communication through low-loss fibers~\cite{kimble_quantum_2008, wehner_quantum_2018}.
Bridging these regimes requires quantum transducers that coherently convert quantum states between different physical carriers~\cite{lauk_perspectives_2020, Lambert2020, zeuthen_figures_2020, han_microwave-optical_2021, awschalom_development_2021, zhao_building_2025}. 
Unlike classical frequency converters, a quantum transducer must operate in the quantum-enabled regime, requiring near-unity conversion efficiency $(\eta \approx 1)$, minimal added noise ($N_\text{add}< 1$ photon), and large bandwidth compatible with qubit coherence. 
Developing a microwave–optical interface that fulfills these criteria simultaneously remains a major challenge~\cite{lauk_perspectives_2020, Lambert2020, zeuthen_figures_2020, han_microwave-optical_2021, awschalom_development_2021, zhao_building_2025}.

Bridging the five orders of magnitude in energy between microwave and optical photons requires strong nonlinear interactions, either directly or involving mediators (e.g., phonons, magnons) or frequency-wave mixing. 
Electro-optic transducers based on the Pockels effect offer a direct and potentially low-noise pathway with MHz-scale bandwidths, but are currently limited in efficiency by the weakness of the underlying electro-optic interaction~\cite{fan_superconducting_2018,mckenna_cryogenic_2020, holzgrafe_cavity_2020,xu_bidirectional_2021,sahu_quantum-enabled_2022}. 
Optomechanical transducers use a mechanical mode to coherently connect microwave and optical fields via electromechanical and optomechanical couplings. 
While internal efficiencies up to $\approx 47\%$ have been reported, their performance is constrained by thermal occupation of the mechanical mediator and a limited bandwidth (often $\text{kHz}-10$ kHz in high-efficiency operation)~\cite{andrews_bidirectional_2014,balram_coherent_2016,brubaker_optomechanical_2022,mirhosseini_superconducting_2020,forsch_microwave--optics_2020,delaney_superconducting-qubit_2022,wang_high-efficiency_2022,van_thiel_optical_2025, zhao_quantum-enabled_2025}.
Other approaches, such as nonlinear frequency mixing in Rydberg atom ensembles~\cite{adwaith_coherent_2019,vogt_efficient_2019, tu_high-efficiency_2022, kumar_quantum-enabled_2023, borowka_continuous_2024} and cavity-enhanced Raman scattering in rare-earth ion ensembles~\cite{zhong_nanophotonic_2015, bartholomew_-chip_2020, rochman_microwave--optical_2023}, have also produced limited success. 
Magnons offer an alternative route to microwave–optical transduction because they can be tuned by static magnetic fields and can support intrinsically broadband dynamics. 
Additionally, low thermal occupancy at higher frequency (GHz) is ideal for low-noise operation.
To date, most magnon-based microwave–optical interfaces have relied on off-resonant magneto-optical effects (e.g., the Faraday effect) where the interaction is intrinsically weak, requiring large device volumes or high-finesse cavities to achieve measurable conversion~\cite{hisatomi_bidirectional_2016, wu_microwave--optics_2025}.

Here, we present an alternative approach that exploits the resonant excitonic susceptibility and coherent magnon–exciton coupling in the van der Waals (vdW) antiferromagnet CrSBr. 
This layered semiconductor combines GHz-frequency magnons (even in the absence of static magnetic field)~\cite{cham_anisotropic_2022, cho_microscopic_2023, Xu2025} and tightly bound large-oscillator-strength excitons~\cite{wilson_interlayer_2021, datta2025magnon}.
It further enables self-hybridized exciton-polaritons in the ultrastrong coupling regime~\cite{dirnberger_magneto-optics_2023, wang_magnetically-dressed_2023, Adak}.
The excitonic resonances are intrinsically linked to the magnetic order, enabling optical access to magnons and magnon-mediated optical nonlinearities~\cite{bae_exciton-coupled_2022, diederich_tunable_2022, sun_dipolar_2024, datta2025magnon}. 
Exploiting this unique light--matter coupling, we demonstrate a new transduction mechanism that combines the broadband, magnetically tunable response of magnons with the strongly coupled optical interface of exciton-polaritons. 
The resulting interaction accesses a resonantly enhanced magneto-optic regime that is qualitatively distinct from off-resonant Faraday-based approaches. 
Combined with its layered, integrable architecture, CrSBr-based magnon-coupled exciton-polaritons offer a scalable and practical route toward efficient and broadband microwave–optical transduction.

\section*{Transduction platform}
\begin{figure}[hbt!]
    \centering
	\includegraphics[width=13.5cm]
    {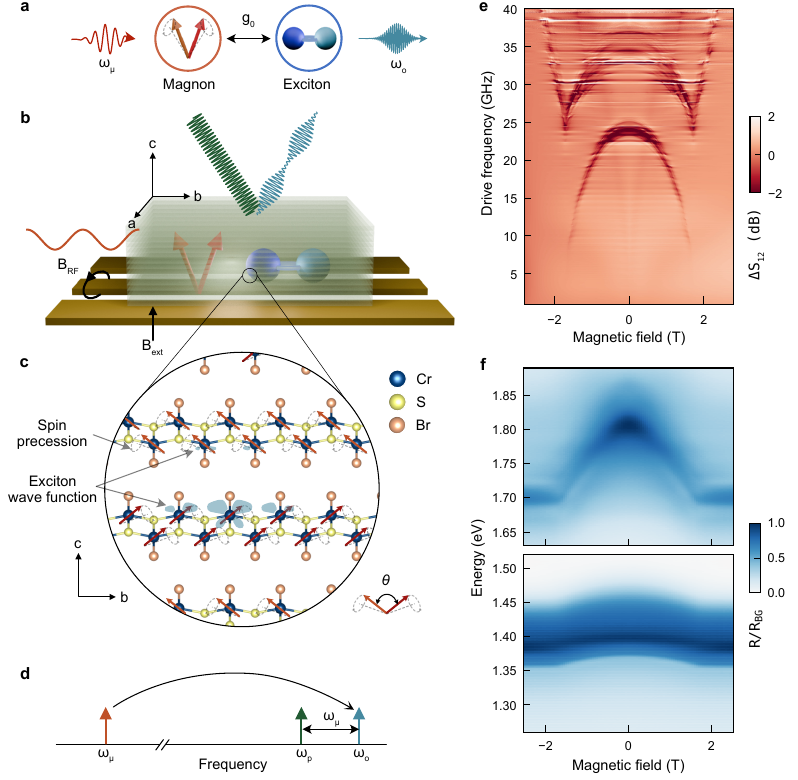}
        \caption{ 
        \textbf{Transduction platform and magneto-optical characterization.}
        \textbf{a},~Schematic of microwave-to-optical conversion mediated by coupling between magnon and exciton. 
        \textbf{b},~Device schematic. A bulk CrSBr crystal is placed on a coplanar waveguide (CPW) inside a cryostat at 2~K. The crystal $b$-axis aligns with the transmission line, with the $c$-axis perpendicular to the CPW plane. Resonant microwave excitation drives magnons, while an optical pump excites excitons. An external, static field $B_\text{ext}$ is applied perpendicular to the CPW. Magnon–exciton coupling produces an optical modulation at magnon frequency. 
        \textbf{c},~Crystal and magnetic structure for $B_\text{ext} < B_\text{sat}$, where $B_\text{sat}$ is the saturation field. Spins in adjacent layers are antiferromagnetically aligned and canted toward the field, setting a canting angle $\theta$. The microwave drive excites a uniform precession of the spins (magnons). Exciton wavefunctions respond to spin alignment, enabling coupling to magnons.
        \textbf{d},~Frequency-domain representation of microwave-to-optical transduction. Optical photons are generated as a sideband at frequency $\omega_o$, offset by the microwave frequency $\omega_{\mu}$ from the optical pump $\omega_p$. 
        \textbf{e},~Microwave transmission change $\Delta S_{21}$ as a function of the magnetic field showing magnon dispersion. Lower and upper branches correspond to optical and acoustic magnon modes, respectively. Above the saturation field ($B_\text{sat} \approx 1.9$~T), a single ferromagnetic resonance branch is observed.
        \textbf{f},~Normalized optical reflectance ($R/R_\text{BG}$) versus the static magnetic field showing exciton dispersion, where $R_\text{BG}$ is the reflectance of the CPW metal plane. Both high- and low-energy excitons show energy redshifts with increasing field magnitude, evidencing coupling between exciton and the magnetic order.
        }
        \label{fig:fig1}
\end{figure}

The transduction platform involves both magnon and exciton resonances in CrSBr (Fig.~\ref{fig:fig1}a). 
As shown in Fig.~\ref{fig:fig1}b, a CrSBr crystal is placed on a coplanar waveguide (CPW) with its crystallographic $b$-axis aligned to the CPW center conductor and the $c$-axis oriented perpendicular to the CPW plane. 
CrSBr is a layered crystal with an orthorhombic structure, in which each layer consists of two staggered CrS planes, sandwiched between Br atoms and layers stack along the $c$-axis (Fig.\ref{fig:fig1}c).
Below the N\'eel temperature (132~K), CrSBr forms an \textit{A}-type antiferromagnet~\cite{guo_chromium_2018, lee_magnetic_2021}.
While strong intralayer exchange aligns Cr spins ferromagnetically along the $b$-axis, a significantly weaker interlayer exchange leads to an antiparallel alignment between successive layers.
An external static magnetic field $B_\text{ext}$ applied along the $c$ direction cants the spins toward the field while preserving the intralayer ferromagnetic order.
The field therefore sets a canting angle $\theta$ between the two magnetic sublattices. 
The canting increases with field until the saturation field $B_\text{sat}$, where the two sublattices become fully aligned ($\theta=0$).
The weak interlayer coupling allows CrSBr to support magnons at GHz frequencies, rather than the THz frequencies typical of traditional antiferromagnets.

CrSBr also hosts multiple excitons, whose energies are strongly coupled to the magnetic configuration~\cite{wilson_interlayer_2021, datta2025magnon}.
In particular, interlayer electron hopping depends on the relative spin alignment: it is spin-forbidden in the antiferromagnetic configuration but becomes progressively allowed as the layers cant toward a ferromagnetic alignment. 
Consequently, both the electronic band structure and exciton energies depend on $\theta$.
As $\theta$ decreases, the exciton wavefunction becomes more delocalized across layers, resulting in a redshift of the exciton resonances.
This magnetic tunability of the excitons is responsible for coherent magnon--exciton coupling~\cite{bae_exciton-coupled_2022, diederich_tunable_2022}, which is the key to microwave--optical transduction, as outlined below.

The transduction occurs through a two-step coherent process.
First, a microwave signal applied through the CPW resonantly drives uniform magnon modes, in which all spins precess about their equilibrium directions set by $B_\text{ext}$. 
In the optical magnon mode, this precession modulates the instantaneous canting angle $\theta(t)$ at the driving frequency.
Second, the time-dependent $\theta(t)$ modulates the exciton energies and, consequently, the refractive index of the crystal near the excitonic resonances.
When a laser slightly detuned from the exciton resonance reflects from the crystal, the resulting reflectance oscillation generates optical sidebands at frequency offsets equal to the magnon frequency (Fig.\ref{fig:fig1}d).

\section*{Characterization of magnons and excitons}
We use a bulk crystal with a nearly rectangular geometry (thickness $\sim 300~\text{\textmu m}$, width $\sim 0.5~\text{mm}$, and length $\sim 2~\text{mm}$) to characterize the excitonic and magnonic responses of CrSBr independently. 
Figure~\ref{fig:fig1}e shows the changes in microwave transmission, obtained by measuring changes in the transmission coefficient $S_{21}$ with a vector network analyzer (Methods). 
Consistent with previous reports~\cite{cham_anisotropic_2022, cho_microscopic_2023, Xu2025}, two magnon branches are observed for $-B_\text{sat} < B_\text{ext} < B_\text{sat}$, with $B_\text{sat}\sim 1.9$~T.
The lower branch corresponds to the optical mode, exhibiting dispersion in which the magnon frequency decreases with increasing static magnetic field magnitude. 
The acoustic branch follows a similar trend but appears at a higher frequency range.
In contrast to the optical mode, the acoustic mode corresponds to in-phase precession of the two sublattice magnetization vectors, leaving $\theta$ and therefore the exciton energy unchanged.
Beyond $B_\text{sat}\sim1.9$~T, where CrSBr becomes ferromagnetic, a single ferromagnetic resonance (FMR) mode emerges, dispersing linearly with $B_\text{ext}$.
Additional splittings appear within these branches, with larger separation at lower magnetic fields, potentially due to nonuniform crystal thickness. 
The split magnon modes have a typical linewidth around $50-300$~MHz, indicating a large bandwidth useful for broadband transduction.
Importantly, due to its large magnetocrystalline anisotropy, CrSBr supports magnon modes even at zero magnetic field (Fig.~\ref{fig:fig1}e).

Figure~\ref{fig:fig1}f presents the normalized reflectivity spectrum measured under white-light illumination from a halogen source on the same crystal. 
The optical response exhibits multiple strong excitonic resonances, with two dominant features: a high-energy exciton ($X_H$) near 1.8~eV and a low-energy exciton ($X_L$) near 1.4~eV. 
Due to strong Fano interference, the reflectance spectrum exhibits peaks rather than dips near the exciton energies. 
Both excitons redshift with increasing $B_\text{ext}$, consistent with earlier studies~\cite{wilson_interlayer_2021, datta2025magnon}. 
The energy shift follows a parabolic dependence $E_X - E_{X,0} \propto B_\text{ext}^2$ up to $B_\text{sat}$, where $E_X$ and $E_{X,0}$ are the exciton energies at finite and zero field, respectively. 
Expressed in terms of $\theta$, the dependence can be written as
$E_X - E_{X,0} = -\Delta_B \cos^2(\theta/2)$,
where $\Delta_B$ denotes the maximum redshift, approximately 120~meV for $X_H$ and 20~meV for $X_L$. 
Notably, transition dipoles of excitons in CrSBr are strongly aligned along the crystallographic $b$-axis, making the excitonic resonances accessible only by light polarized along the crystallographic $b$-axis.

\section*{Magnon–exciton coupling and transduction measurement}
\begin{figure}[hbt!]
    \centering
	\includegraphics[width=17.5cm]{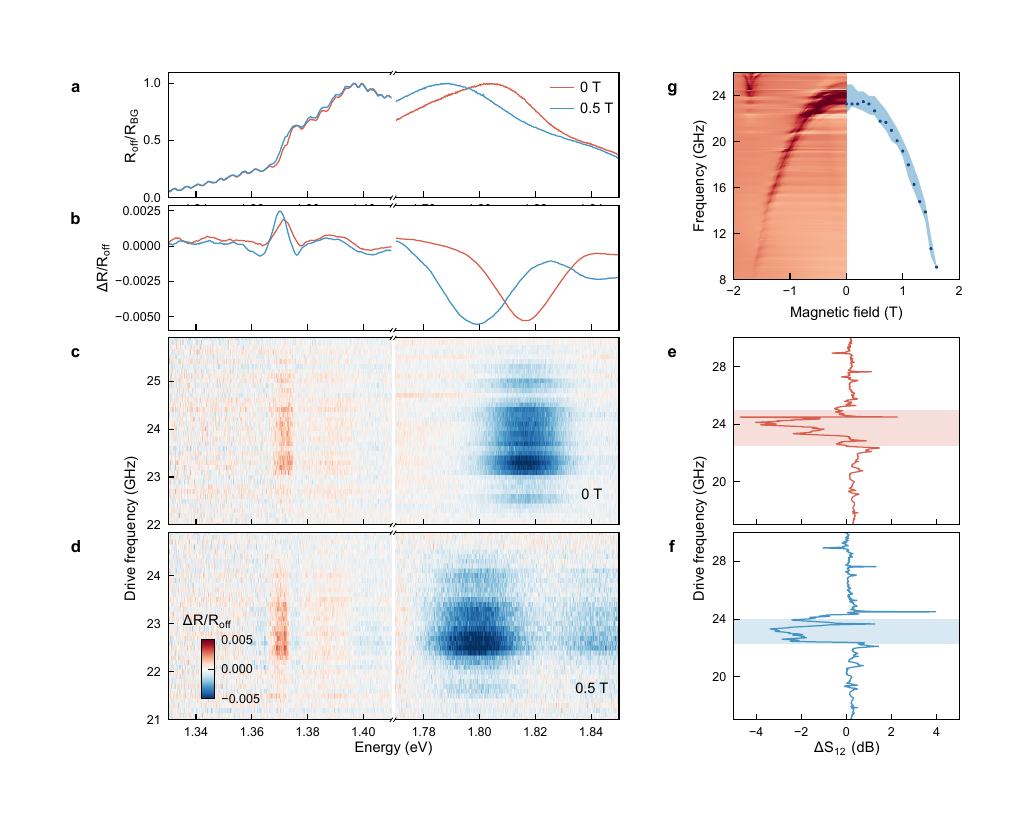}
        \caption{
        \textbf{Magnon-induced reflectance change.}
        \textbf{a},~Normalized optical reflectance  ($R_\text{OFF}/R_{BG}$) versus probe photon energy at $B_\text{ext}=0$~T and $0.5$~T, with the microwave drive off.
        \textbf{b},~Microwave-induced reflectance change, $(\Delta R/R_\text{OFF})= (R_\text{ON}-R_\text{OFF})/R_\text{OFF}$, for both fields. Here, $R_\text{ON}$ ($R_\text{OFF}$) is the reflectance with microwave drive on (off). 
        The microwave drive frequency is 23.2~GHz and 22.5~GHz at $B_\text{ext}=0$~T and $0.5$~T, respectively. 
        The response is strongest at the exciton resonances.
        \textbf{c, d},~Maps of $(\Delta R/R_\text{OFF})$ versus probe energy and microwave drive frequency at $B_\text{ext}=0$~T (\textbf{c}) and $B_\text{ext}=0.5$~T (\textbf{d}).
        \textbf{e, f},~Line cuts of $\Delta S_{21}$ from Fig.~1\textbf{e} at $B_\text{ext}=0$~T (\textbf{e}) and $B_\text{ext}=0.5$~T (\textbf{f}). The highlighted frequency window corresponds to the modulation band in (\textbf{c, d}).
        \textbf{g},(right, positive B)~Field dependence of the drive frequency yielding the maximal reflectance change (peak of $\Delta R/R_\text{OFF}$). The shaded band indicates the frequency range over which an appreciable reflectance change is observed ($\Delta R/R_\text{OFF}>0.0015$). The resulting dispersion agrees with the magnon mode dispersion extracted from microwave spectroscopy (Fig.~1\textbf{e}, left, negative B).
        }
        \label{fig:fig2}
\end{figure}

We now explore magnon–exciton coupling in CrSBr and the resulting microwave-to-optical transduction properties using optical reflectance spectroscopy under microwave excitation. 
Figure~\ref{fig:fig2}a presents the normalized reflectance measured near the two excitons at magnetic fields $B_\text{ext}=0$ and 0.5~T, with microwave drive off. 
The magnetic field induces redshifts in both the exciton energies.
The small oscillations in the spectra around $X_L$ are measurement artifacts arising from the etalon effect of the spectrometer's CCD camera.
We then drive microwave signals oscillating near the magnon resonance frequencies at the applied static magnetic field, as determined in Fig.~\ref{fig:fig1}e.
Figure~\ref{fig:fig2}b shows the relative change in reflectance between microwave drive on and off conditions.
A clear microwave-induced change in reflectivity ($\Delta R/R_\text{off} \equiv (R_\text{on} - R_\text{off})/R_\text{off}$) is observed around the exciton energies. 
Here, $R_\text{on}$ and $R_\text{off}$ are the optical reflectance from the crystal with microwave drive on and off, respectively.
Figures~\ref{fig:fig2}c and \ref{fig:fig2}d show maps of the reflectivity change as a function of optical probe energy and microwave drive frequency without static magnetic field and for a 0.5 T static magnetic field.
Figures~\ref{fig:fig2}e and \ref{fig:fig2}f display the corresponding microwave transmission spectra, obtained from the data presented in Fig.~\ref{fig:fig1}e. 
A pronounced change in $\Delta R/R_\text{off}$ is observed precisely at the intersection of the magnon frequency and exciton energy for each field, confirming magnon–exciton coupling.
We further observe that the change in optical reflectivity varies linearly with the drive microwave power and remains robust up to $\sim$20~K (Supplementary Information). 
The reduced signal at higher temperature could be due to deviation from resonance conditions as well as increased microwave and excitonic dissipation.

We performed similar measurements over a range of magnetic fields and summarized the results in Fig.~\ref{fig:fig2}g.
The right panel of Fig.~\ref{fig:fig2}g plots the microwave drive frequency corresponding to peak reflectivity change as a function of $B_\text{ext}$.
The shaded band depicts the frequency range over which a significant reflectivity change is observed ($\Delta R/R_\text{off}>0.0015$).
The left panel shows the magnon dispersion reproduced from Fig.~\ref{fig:fig1}e. 
The correspondence between the two panels validates the strong magnon–exciton coupling, as well as optical detection of magnons.
These observations point to a time-averaged modulation of the reflectivity induced by microwave-driven magnon precession, evidencing the underlying microwave--optical transduction process.

\section*{Homodyne detection of coherent conversion}
\begin{figure}[hbt!]
    \centering
	\includegraphics[width=16cm]{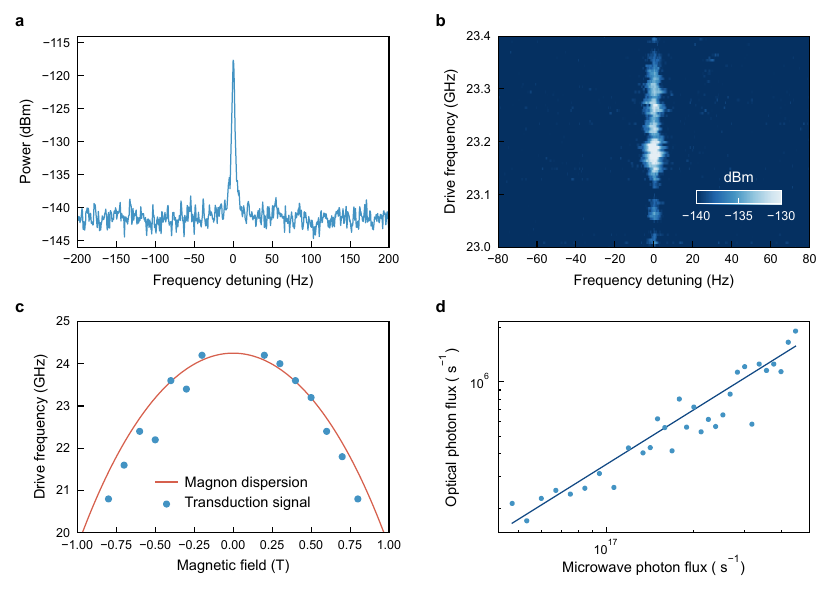}
        \caption{
        \textbf{Homodyne detection of coherent microwave-to-optical conversion.}
        \textbf{a},~Microwave power of the homodyne signal measured on a spectrum analyzer as a function of probe-frequency detuning from the driven magnon frequency. A peak at zero detuning indicates coherent optical modulation at the magnon frequency.
        \textbf{b},~Two-dimensional map of microwave power versus microwave drive frequency and detuning. The modulation persists over a frequency range of $\sim$300~MHz.
        \textbf{c},~Magnetic-field dependence of the drive frequency for which the maximal homodyne signal is observed, consistent with the magnon dispersion from microwave spectroscopy (Fig.~1\textbf{e}).
        \textbf{d},~Converted optical sideband photon flux versus incident microwave photon flux. The solid line is a linear fit. Measurements in Fig.\ref{fig:fig3}\textbf{a}, \textbf{b}, and \textbf{d} were performed with $B_\text{ext}=0.5$~T.
        }
        \label{fig:fig3}
\end{figure}

To validate the coherent microwave-to-optical energy conversion, we employ an amplitude-modulated homodyne detection scheme that enables frequency-domain visualization of the transduction process (Methods). 
A continuous-wave (CW) diode laser with energy 1.8~eV (near $X_H$) is split into a signal arm and a local oscillator (LO). 
The signal beam reflects from the CrSBr crystal, where it acquires sidebands from the magnon-modulated reflectance, before recombining with the LO that is amplitude modulated by a mechanical chopper. 
The resulting homodyne interference yields a signal proportional to the transduced optical field amplitude scaled by the LO power. 
The frequency of the homodyne signal is measured by a fast photodiode and an~RF spectrum analyzer.
Figure~\ref{fig:fig3}a shows a measurement at a static field of $B_{\text{ext}} = 0.5$~T, where the exciton energy is tuned into near-resonance with the laser.
We observe a robust transduction signal at zero frequency detuning from the magnon sideband.
The narrow peak width reflects the high spectral purity of the microwave drive.
Consistent with the excitonic transition dipole orientation in CrSBr, the signal is prominent for $b$-polarized incident light but vanishes into the noise floor for $a$-polarized light (Supplementary Information).

We can infer the transduction bandwidth by recording the homodyne signal versus drive frequency (Fig.~\ref{fig:fig3}b). 
The response extends over $\sim$300~MHz, consistent with the magnon linewidth observed in microwave spectroscopy (Fig.~\ref{fig:fig1}e).
Figure~\ref{fig:fig3}c demonstrates the tunability of this platform by tracking the frequency of the transduction signal across a wide range of static magnetic fields (details in Supplementary Information).
The field dependence of the drive frequency for transduction follows the magnon dispersion, as highlighted by the overlaid curve in Fig.~\ref{fig:fig3}c.

To quantify the transduction efficiency, we convert the measured homodyne sideband into photon fluxes and relate it to the absorbed microwave drive.
We obtain the transduced optical photons by noting various coupling losses in the optical path, photodetector responsivity, and RF amplification.
The input microwave photon flux is determined from the RF power and S parameter measurements.
Since the dissipation pathways inside the cryostat cannot be fully disentangled, the measured transmission and reflection provide a conservative upper bound on the microwave power absorbed by the crystal, and hence on the absorbed microwave photon flux.
In Fig.~\ref{fig:fig3}d, the transduced optical photon flux ($N_{o,\text{out}}$) is plotted against the upper bound of input microwave photon flux ($N_{\mu,\text{in}}$).
At low microwave power we observe a linear dependence, consistent with direct correspondence between input microwave photon and output optical photon numbers. 
From the slope of the fitting line, we extract a lower limit of the transduction efficiency, $\eta\equiv \frac{N_{o,\text{out}}}{N_{\mu,\text{in}}}=3.3\times 10^{-12}$.
A modest efficiency is expected for the present bulk implementation: the microwave drive excites magnons over the full crystal volume, whereas the optical probe samples only a micron-scale region, and the interaction is not enhanced by optical and/or microwave photonic resonances.

\section*{Physical insights and polariton engineering}
To understand the fundamental limits and ultimate potential of this platform, we analyze the transduction mechanism. 
The coupled magnon--exciton system is described using the Hamiltonian $H = \hbar \omega_{x}(\theta) x^\dagger x + \hbar \omega_m m^\dagger m$, where $x^\dagger$ ($m^\dagger$) and $\omega_x$ ($\omega_m$) are the creation operators and resonant frequencies for excitons (magnons), respectively.
The coupling is rooted in the dependence of the exciton resonance on the canting angle $\theta$, which is modulated by the magnon mode.
For small excursions about the static canting angle $\theta_0$, the Hamiltonian can be cast into the linearized form, $H_{me}\simeq \hbar \omega_{x,0} x^\dagger x + \hbar \omega_m m^\dagger m + g_0(m + m^\dagger)x^\dagger x$.
Here, $\omega_{x,0}$ is the zero field exciton frequency and $g_0$ is the single-particle magnon–exciton coupling rate defined as $g_0 \equiv (\partial \omega_x/\partial \theta)_{\theta_0}\, \theta_\text{ZPF}$. 
The zero-point fluctuation of the canting angle is given by $\theta_\text{ZPF} \approx \sqrt{V_\text{unit}/(3V_\text{mag})}$, where $V_\text{unit}$ is the unit-cell volume of CrSBr, and $V_\text{mag}$ is the effective magnetic volume.
This expression reveals an important scaling law: the coupling is inversely proportional to $\sqrt{V_\text{mag}}$.
Further, $g_0$ is tuned by the magnetic field through $(\partial\omega_x/\partial\theta)_{\theta_0}$ and is maximum near $\theta_0\simeq \pi/2$. 
The high-energy exciton $X_H$ exhibits a larger field-induced energy shift than $X_L$, leading to a stronger $g_0$.

In our bulk crystals, the substantial $V_\text{mag}$ results in a small precession angle per magnon, yielding a single-particle coupling strength of $g_0/2\pi \approx 5.5$~kHz for the $X_H$ exciton. 
For perspective, this value is approximately three orders of magnitude larger than the magneto-optical coupling typically observed in YIG~\cite{han_microwave-optical_2021}. 
Unlike 3D ferrimagnets, the 2D nature of CrSBr allows for extreme confinement. 
Reducing $V_{\text{mag}}$ to the few-layer limit increases $\theta_{\text{ZPF}}$, potentially pushing $g_0$ into the MHz regime, surpassing the rates of most current optomechanical and electro-optic converters (see Supplementary Information).
To evaluate the transduction efficiency, we define the cooperativity as $C = 4|G|^2 / (\gamma_m \gamma_x)$, where $\gamma_m$ and $\gamma_x$ represent the magnon and exciton linewidths, respectively. 
Here, $G = g_0 \sqrt{n_x}$ denotes the pump-enhanced coupling rate, where $n_x$ is the exciton population. 
Despite the high $g_0$, $C$ is constrained by the relatively broad excitonic linewidths. 
Consequently, the single-exciton cooperativity ($n_x = 1$) remains modest in a bulk crystal $\sim 10^{-14}$.
This value is consistent with our estimation from the experiment corresponding to an exciton population of $\sim 10^{2}$.

\begin{figure}[hbt!]
    \centering
	\includegraphics[width=8.5cm]{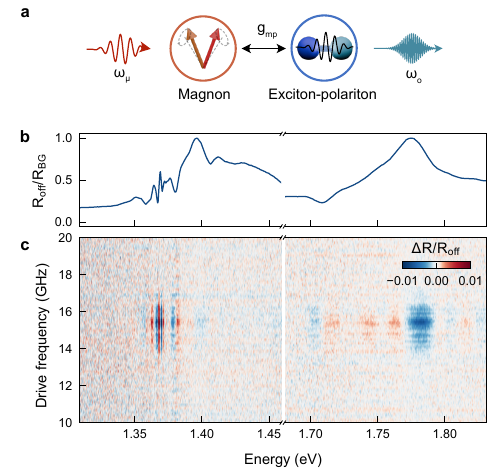}
        \caption{
        \textbf{Coupling between magnons and exciton-polaritons.}
        \textbf{a},~Schematic of microwave-to-optical conversion mediated by coupling between magnon and exciton-polariton.
        \textbf{b},~Normalized reflectance $R/R_\mathrm{BG}$ versus probe photon energy, taken at $B_\text{ext}=-1.3$~T, showing multiple polariton resonances.
        \textbf{c},~Microwave-induced differential reflectance $\Delta R/R_\mathrm{OFF}$  at $B_\text{ext}=-1.3$~T as a function of microwave drive frequency and probe photon energy, revealing multiple polariton-assisted transduction features.
        }
        \label{fig:fig4}
\end{figure}

The inherent dissipative loss due to excitonic absorption can be further addressed by harnessing exciton-polaritons, hybrid quasiparticles formed by the strong coupling of excitons and cavity photons. 
Exciton-polaritons in CrSBr also couple to magnons, making them suitable for transduction (Fig.~\ref{fig:fig4}a).
The large index contrast between air and CrSBr forms an effective Fabry-P\'erot cavity, producing multiple self-hybridized exciton-polariton resonances near both $X_H$ and $X_L$ (Fig.~\ref{fig:fig4}b)\cite{dirnberger_magneto-optics_2023,Adak}.
As shown in Fig.~\ref{fig:fig4}c, the microwave-induced reflectivity changes across multiple polariton branches, confirming that these hybrid states inherit the magnon--exciton coupling and enable microwave--optical transduction.
A polariton mode $p$ can be written as a coherent superposition of an exciton and a photon,
$p = \chi_x x + \chi_{\rm ph} a$, with Hopfield coefficients $\chi_x$ and $\chi_{\rm ph}$ satisfying $|\chi_x|^2+|\chi_{\rm ph}|^2=1$.
The magnon--polariton coupling scales as $g_{mp}\approx |\chi_x|\, g_0$, while the polariton linewidth $\gamma_p$ can be reduced by borrowing photonic character (limited by the optical loss rate of the corresponding photonic mode). 
Consequently, the cooperativity $C_{mp} = 4g_{mp}^2 n_p / (\gamma_m \gamma_p)$ remains relatively robust. 
Here, $n_p$ is the exciton-polariton population.
Thus, polariton engineering allows one to mitigate dissipative loss without sacrificing transduction efficiency, which could be beneficial to minimize added noise in the transduction process.
Furthermore, the hybridization widens the practical optical conversion window without requiring operation exactly on the bare exciton resonance.

\section*{Discussion and outlook}
The path toward near-unity internal efficiency relies on two strategies: volume confinement and pump enhancement. 
Reducing $V_{\rm mag}$ toward the optical mode volume increases $\theta_{\rm ZPF}$ and therefore boosts $g_0$ dramatically, while a microwave resonator can simultaneously enhance the microwave photon--magnon coupling to compensate for the reduced magnetic volume. 
Recent demonstrations of strong microwave photon–magnon coupling in exfoliated $\text{CrSBr}$ flakes underscore the viability of this approach~\cite{tang_coherent_2025}. 
Crucially, the layered vdW nature of $\text{CrSBr}$ ensures that both magnon and exciton linewidths remain relatively unchanged even in the few-layer limit. 
Using experimentally reported linewidths for these low-dimensional structures, we estimate a single-exciton cooperativity of $C \approx 10^{-9}$ in typical $\text{CrSBr}$ flakes, reaching $10^{-4}$ in idealized bilayer implementations (see Supplementary Information).
These values position $\text{CrSBr}$ in a regime where combined cavity- and pump-enhancement can realistically drive the cooperativity toward the efficiency limit.
Given the high oscillator strength of CrSBr excitons and the ability to integrate these vdW flakes into high-finesse microcavities, achieving the high cooperativity required for quantum-coherent microwave-to-optical transduction is a realistic near-term goal.

\section*{Acknowledgments}
\subsubsection*{Funding:}
Work at City College was primarily supported by the U.S. Department of Energy (DOE), Office of Science, Basic Energy Sciences (BES), under Award DE-SC0025302 (microwave spectroscopy), DARPA grant HR0011-25-3-0107 (design  and fabrication of microwave waveguides), and the Gordon and Betty Moore Foundation grant 12764 (optical spectroscopy). 
Z.S. was supported by project LUAUS25268 from Ministry of Education Youth and Sports (MEYS), ERC-CZ program (project LL2101) from Ministry of Education Youth and Sports (MEYS) and by the project Advanced Functional Nanorobots (reg. No. CZ.02.1.01/0.0/0.0/15\_003/0000444 financed by the EFRR) (growth and synthesis). 
K.M. were supported from the grant of Specific university research – grant No A1\_FCHT\_2025\_013 (growth and synthesis).
Synthesis work at Columbia was supported by the Materials Science and Engineering Research Center (MRSEC) on Precision Assembly of Quantum Materials (PAQM) through NSF award DMR-2011738.
A.K. was supported by the German Research Foundation (DFG) via Spin+X TRR 173-268565370, project A13.
A.A. acknowledges support from the Office of Naval Research with grant No. N000142612008.

We acknowledge Agneya V Dharmapalan and Ananthu Mahendranath for help with experiments; Shaedil U Dider and Akshaj Arora for help in PCB fabrication; Supriya Mandal, Rohin Verma, and Matt Shmukler for discussions about microwave measurements.

\subsubsection*{Author Contributions Statement:}
P.C.A. and V.M.M. conceived the project. P.C.A. designed the experiments, led the measurements, and developed the analysis framework. 
K.M., A.R, X.R. and Z.S. synthesized the CrSBr crystals. 
P.C.A. and I.M. performed device fabrication. 
I.M. carried out data analysis and visualization under P.C.A.’s guidance. P.C.A. developed the theoretical model with input from C.H-H. S.P., I.M., and B.D. assisted with experiments. V.K. contributed to the microwave spectroscopy setup under the supervision of A.T. and A.A. T.Z. and A.K. provided insights into the transduction mechanism. P.C.A. wrote the manuscript with input from I.M. and V.M.M. All authors commented on the manuscript. V.M.M. supervised the project.

\subsubsection*{Competing Interests Statement:}
P.C.A and V.M.M have a patent pending. The remaining authors declare no competing interests.

\clearpage

\clearpage

\subsection*{Materials and Methods}
\subsubsection*{Device fabrication}
CrSBr bulk crystals were grown using the chemical vapor transport method.~\cite{klein_control_2022} CPWs were designed in COMSOL and fabricated using a LPKF Protomat S104 and ProtoLaser U4 using  17.5~\textmu m thick Cu on 1.5~mm thick FR4. 

\subsubsection*{Microwave spectroscopy}
All measurements were conducted on a CrSBr sample mounted on a CPW with the $b$-axis aligned with the center conductor. This device was then placed in a closed cycle cryostat (Quantum Design Opticool) with the tunable static magnetic field (-7 to 7 T) perpendicular to the plane of the CPW aligned with the $c$-axis of the crystal. All data was taken at a base temperature of $\sim$2 K unless otherwise mentioned. CW Microwave transmission measurements were taken using a PNA. The change in microwave transmission $\Delta S_{12}$ was calculated by subtracting magnetic field averaged background to remove frequency dependent streaking.

\subsubsection*{Optical reflectance measurement}
Reflectance spectra were obtained using a spectrally broadband tungsten--halogen lamp that was linearly polarized to align with the $b$-axis of the CrSBr crystal. The light was then focused onto a 6~\textmu m spot on the sample with a $100\times$ objective (NA $=0.8$) and the reflectance measured with a grating spectrometer. Optical reflectance data ($R/R_\text{BG}$) was normalized with background taken on the copper surface of the CPW. To find the microwave induced differential reflectivity $(R_\text{ON}-R_\text{OFF})/R_\text{OFF}$, we measured $R_\text{ON}$ by driving the CrSBr with a 13 dBm CW microwave source connected to the cryostat's RF ports using coaxial cables. The polariton differential reflectance measurements were driven with a 10 dBm signal from the PNA. To reduce noise, we applied a Savitzky-Golay filter to the differential reflectivity along the energy axis.


\subsubsection*{Homodyne measurement}
The homodyne setup used a 688 nm CW diode laser split into a signal and a local oscillator (LO) path using a 30:70 beam splitter. The signal path was reflected off CrSBr, which was driven by the RF signal generator. The LO path was modulated by a $\sim$677 Hz chopper and passed through a neutral density (ND) filter to control LO power. Each path was coupled to single-mode optical fiber and then combined in a polarization-maintaining $2\times2$ 50:50 coupler. One output then passed through a fast photodetector and a low noise amplifier before it was measured with a spectrum analyzer.
The spectrum analyzer was synchronized with the RF signal generator. Unless otherwise noted, the homodyne measurement was taken at $B = 0.5$~T and the microwave drive frequency was 23.181~GHz.


\clearpage

\end{document}